\documentclass[10pt,twocolumn,letterpaper]{article}

\usepackage{cvpr}
\usepackage{times}
\usepackage{epsfig}
\usepackage{graphicx}
\usepackage{amsmath}
\usepackage{amssymb}
\usepackage{amsfonts}
\usepackage{dsfont}
\usepackage{array}

\usepackage[pagebackref=true,breaklinks=true,letterpaper=true,colorlinks,bookmarks=false]{hyperref}

\cvprfinalcopy 

\ifcvprfinal\fi
\begin{document}

\title{Augmenting Colonoscopy using Extended and Directional CycleGAN \\ for Lossy Image Translation}

\author{Shawn Mathew$^{1*}$ \and Saad Nadeem$^2${\footnote{Equal Contribution}} \and Sruti Kumari$^1$ \and Arie Kaufman{$^1$}\\ \and \and
$^1${\small Stony Brook University} \and $^2${\small Memorial Sloan Kettering Cancer Center}\\ \and
{\tt\small \{shawmathew,srkumari,ari\}@cs.stonybrook.edu} \and {\tt\small nadeems@mskcc.org}
}

\maketitle

\begin{abstract}
\let\thefootnote\relax\footnotetext{\hspace{-3mm}* Equal Contribution}Colorectal cancer screening modalities, such as optical colonoscopy (OC) and virtual colonoscopy (VC), are critical for diagnosing and ultimately removing polyps (precursors of colon cancer). The non-invasive VC is normally used to inspect a 3D reconstructed colon (from CT scans) for polyps and if found, the OC procedure is performed to physically traverse the colon via endoscope and remove these polyps. In this paper, we present a deep learning framework, Extended and Directional CycleGAN, for lossy unpaired image-to-image translation between OC and VC to augment OC video sequences with scale-consistent depth information from VC, and augment VC with patient-specific textures, color and specular highlights from OC (e.g, for realistic polyp synthesis). Both OC and VC contain structural information, but it is obscured in OC by additional patient-specific texture and specular highlights, hence making the translation from OC to VC lossy. The existing CycleGAN approaches do not handle lossy transformations. To address this shortcoming, we introduce an extended cycle consistency loss, which compares the geometric structures from OC in the VC domain. This loss removes the need for the CycleGAN to embed OC information in the VC domain. To handle a stronger removal of the textures and lighting, a Directional Discriminator is introduced to differentiate the direction of translation (by creating paired information for the discriminator), as opposed to the standard CycleGAN which is direction-agnostic. Combining the extended cycle consistency loss and the Directional Discriminator, we show state-of-the-art results on scale-consistent depth inference for phantom, textured VC and for real polyp and normal colon video sequences. We also present results for realistic pendunculated and flat polyp synthesis from bumps introduced in 3D VC models. 

\end{abstract}

\section{Introduction}

Colon cancer is one of the most commonly diagnosed cancers with 1.8 million new cases (and subsequent 750,000 deaths) reported worldwide every year \cite{bray2018global}. Optical colonoscopy (OC) is the most prevalent colon cancer screening procedure. In this invasive procedure, polyps (precursors of colon cancer) can be found and removed using an endoscope. In contrast, virtual colonoscopy (VC) is a non-invasive screening procedure where the colon is 3D reconstructed from computed tomography (CT) scans and inspected for polyps with a virtual flythrough (simulating the OC endoscope traversal). Due to its non-invasive, inexpensive, and low-risk (no sedation required) nature, VC is becoming a commonplace tool for colon cancer screening. In fact, the US Multi-Society Task Force on Colorectal Cancer recommends VC screenings every 5 years and OC every 10 years for average-risk patients above the age of 50 \cite{rex2017colorectal}.

\begin{figure*}[t!]
\begin{center}
\footnotesize
\setlength{\tabcolsep}{6pt}

\begin{tabular}{ccc|ccc}
&  & Histogram-Equalized &  &  & Histogram-Equalized\\
OC      &    Synthetic VC    & Output   &  OC    &  Synthetic VC      & Output\\
\includegraphics[width=0.11\textwidth]{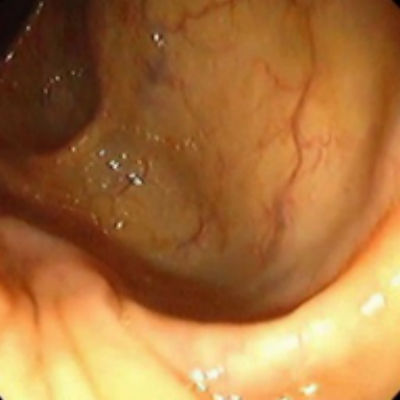}&
\includegraphics[width=0.11\textwidth]{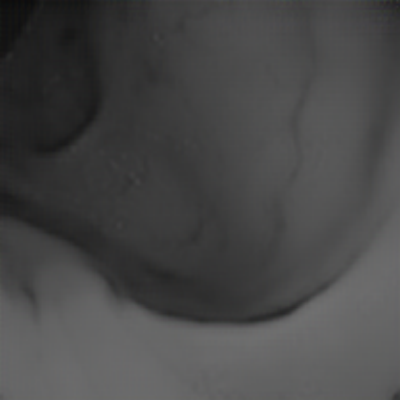}&
\includegraphics[width=0.11\textwidth]{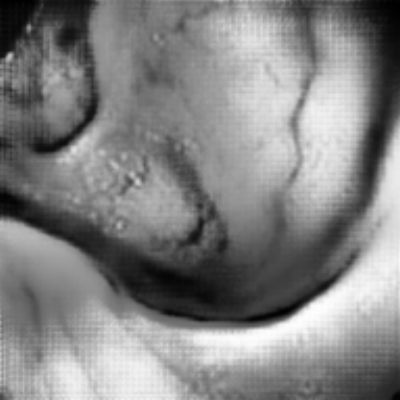}&

\includegraphics[width=0.11\textwidth]{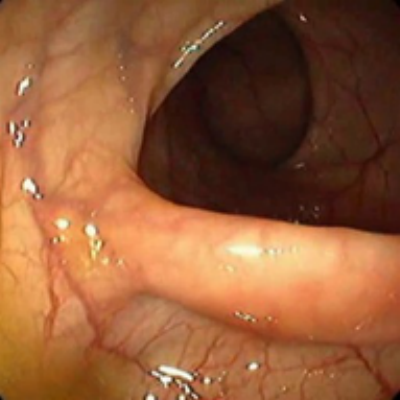}&
\includegraphics[width=0.11\textwidth]{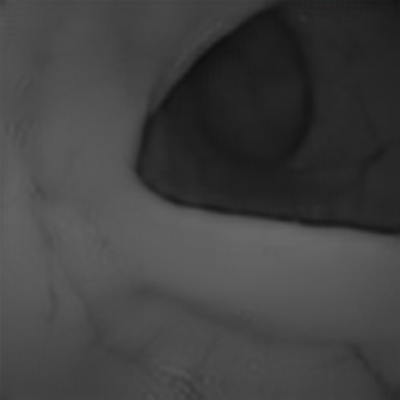}&
\includegraphics[width=0.11\textwidth]{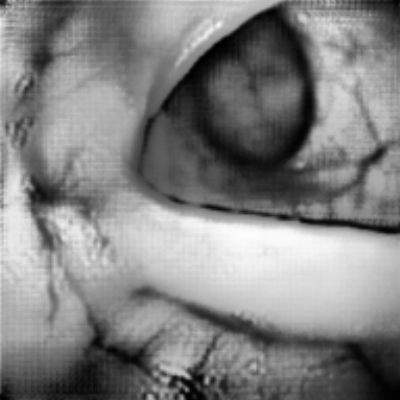}\\

\end{tabular}
\caption{Two examples of standard CycleGAN lossy transformation problem \cite{chu2017cyclegan}. In OC to VC translation, standard CycleGAN stores the textures and specular reflections in the VC domain as depicted in the histogram-equalized output.}
\label{fig:CLAHE}
\end{center}
\end{figure*}

Both VC and OC provide complementary information. OC endoscope videos are comprised of individual frames capturing complex real-time dynamics of the colon with important texture information (e.g., veins, blood clots, stool, etc). VC, on the other hand, provides complete 3D geometric information of the colon including polyps. This complementary nature of OC and VC motivates our current work to find ways of translating information between these two modalities. The geometric information from VC images can aid in 3D reconstruction and surface coverage (percentage of colon inspected) during the OC procedure; lower the surface coverage higher the polyp miss rate. Inferring scale-consistent depth maps for given OC video sequences enables 3D reconstruction through established simultaneous localization and mapping (SLAM) algorithms \cite{schops2019surfelmeshing,whelan2015elasticfusion}, which can help deduce the surface coverage during OC. Augmenting VC with texture and specular highlights from OC can be used to generate realistic virtual training simulators for gastroenterologists as well as realistic polyps. Shin et al. \cite{shin2018abnormal} have presented a method to produce polyps from edge maps and binary polyp masks. This generates realistic polyps, but the 3D shape and endoscope orientation are hard to control making it difficult to produce specific polyp shapes, for example, flat polyps. VC to OC translation, in our context, provides full control over the 3D shape and endoscope orientation making it easy to generate pendunculated and flat polyps.

The task of translating between OC and VC can be generalized to image-to-image domain translation. Since there is no ground truth paired data for OC and VC, CycleGAN \cite{zhu2017unpaired} is suited to this problem, but it cannot handle lossy transformations, for example, between VC (structure) and OC (structure + color + texture + specular highlights), as shown by Chu et al. \cite{chu2017cyclegan}. Porav et al. \cite{porav2019reducing} have presented a method to handle the lossy CycleGAN translation by adding a denoiser to reduce high frequencies with low amplitudes. As seen in Figure \ref{fig:CLAHE}, specular highlights and textures are not embedded as high frequency/low amplitude signals, hence the denoiser will not help in our context.

Thus, we introduce a novel extended cycle consistency loss for lossy image domain translation. This frees the network from needing to hide information in the lossy domain by replacing OC comparisons with VC comparisons. Stronger removal of these specular reflections and textures are handled via a Directional Discriminator that differentiates the direction of translation as opposed to the standard CycleGAN which is direction-agnostic. This Directional Discriminator acts like a discriminator in a conditional GAN and deals with paired data thus giving the network, as a whole, a better understanding of the relationship between the two domains.

The contributions of this work are as follows:
\begin{enumerate}
\itemsep 0em 
    \item A lossy image-to-image translation model via a novel extended cycle consistency loss to remove texture, color and specular highlights from VC.
    \item A Directional Discriminator to create a stronger link between OC and VC for removing remaining textures and lighting.
    \item The same framework can synthesize realistic OC (flat and pendunculated) images.
    \item  Scale-consistent depth inference from OC video sequences.
\end{enumerate}

\section{Related Work}
\noindent
\textbf{Generative Adversarial Networks:} GANs \cite{goodfellow2014generative} introduced the concept of adversarial learning and have shown promising results in image generation, segmentation \cite{long2015fully}, super resolution \cite{ledig2017photo}, video prediction \cite{mathieu2015deep} and more. The idea behind GANs revolves around two networks playing a game against each other.

\begin{figure*}[t!]
\centering
\includegraphics[ width=0.85\textwidth]{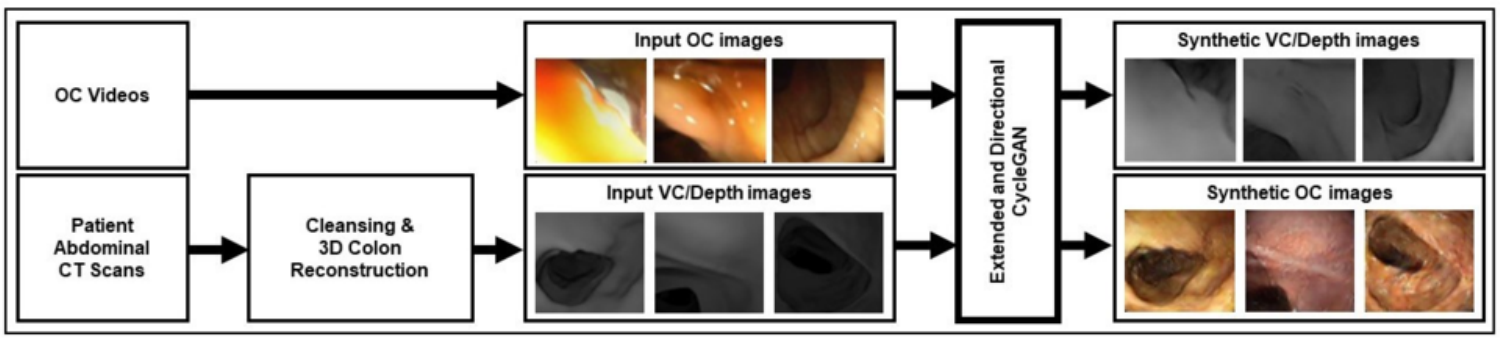}
\caption{Pipeline for generating realistic VC and OC images from their counterpart. OC and VC images are extracted from videos. VC videos are created from reconstructing CT scans and then rendering a flythrough of the colon. This data is passed into the generators of the Extended and Directional CycleGAN and produce VC and OC images.} 
\label{fig:pipeline}
\end{figure*}

\noindent
\textbf{Image-to-Image Translation:} This task maps an image in one domain to another. OC and VC image translation, in our context, can be reframed as an image-to-image translation problem. The pix2pix network is a deep learning model that solves this problem using a conditional GAN with an additional L1 loss \cite{isola2017image}. This model requires paired ground truth data from two given domains, which is not available in our context.

Recent deep learning approaches that tackle unpaired image-to-image domain translation include CycleGAN \cite{zhu2017unpaired} and similar approaches \cite{kim2017learning,yi2017dualgan}. In this paper, we modify CycleGAN for unpaired lossy image-to-image translation between OC and VC, and further alter it to create a stronger link between the two input domains. CycleGANs, have been shown to hallucinate features \cite{shin2018abnormal}, which is problematic if used directly for patient diagnosis. We, however, use it as an add-on to the real data rather than for diagnostic purposes.

CycleGANs, when dealing with lossy image translations, tend to hide information in the lossy images. The cycle consistency loss requires the network to embed extra information in the lossy domain, in order to reconstruct the image \cite{chu2017cyclegan}. Porav et al. \cite{porav2019reducing} have proposed a possible solution to the lossy domain translation by adding a denoiser to reduce high frequencies with low amplitudes. In our case, the network simply tries to blend in texture and lighting artifacts with the colon wall, so their method is not helpful.

Mirza et al. \cite{mirza2014conditional} have introduced the idea of conditioning the output of the generator with all or part of the input. This extra information is passed to the discriminator and provides a stronger link between the input and the output. Conditional CycleGAN \cite{lu2018attribute} employs this same concept, where a label or another image are used to drive the direction of translation. In other words, the CycleGAN allows for extra input to drive the translation but requires ground truth pair between the label and the input. Our Directional Discriminator is similar to conditional GANs, but unlike conditional GANs does not require the ground truth labels and input.

Donahue et al. \cite{donahue2016adversarial} and Dumoulin et al. \cite{dumoulin2016adversarially} have presented approaches that are similar to ours as they use paired input and output of two networks to train a single discriminator, but instead of pairing images (like in our case), they pair latent vectors and images. As shown by Zhu et al. \cite{zhu2017unpaired}, these approaches did not work well by themselves in the image-to-image domain translation task and resulted in heavy artifacts and unrealistic images. More recently, Pajot et al. \cite{pajot2018unsupervised} have discussed a similar extended cycle consistency loss to ours for reconstructing noisy images. We differ from their method as we only take one step forward in the cycle to allow for a one-to-many image translation (requirement for our application), rather than two steps forward in their case.\\

\begin{figure}[t!]
\centering
\includegraphics[width=0.4\textwidth]{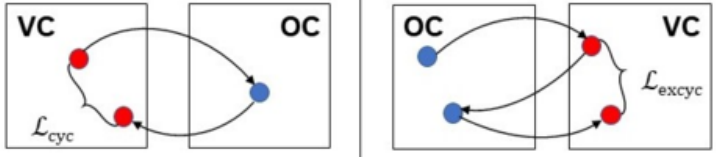}
\caption{The image on the left depicts the cycle consistency loss used for VC to OC translation from Zhu et al's CycleGAN \cite{zhu2017unpaired}. The image on the right shows the extended cycle consistency loss that we used for OC to VC translation.}
\label{fig:difference}
\end{figure}

\begin{figure}[t!]
\centering
\setlength{\tabcolsep}{1pt}
\begin{tabular}{ccccc}
Input $OC$  & $VC_{syn}$& $OC_{rec}$ & $VC_{syn\_rec}$ \\
\includegraphics[width=0.1\textwidth]{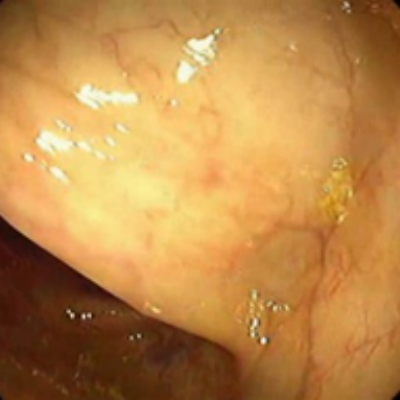}&
\includegraphics[width=0.1\textwidth]{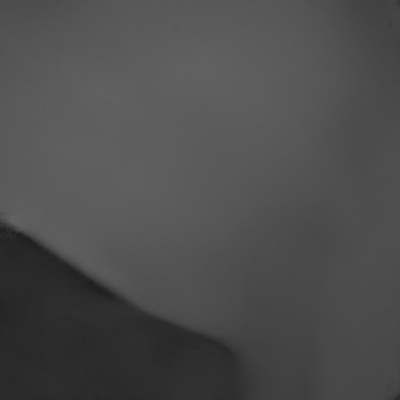}&
\includegraphics[width=0.1\textwidth]{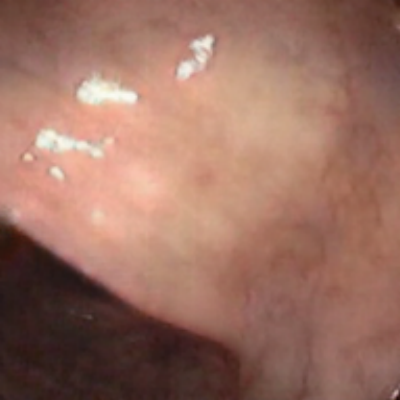}&
\includegraphics[width=0.1\textwidth]{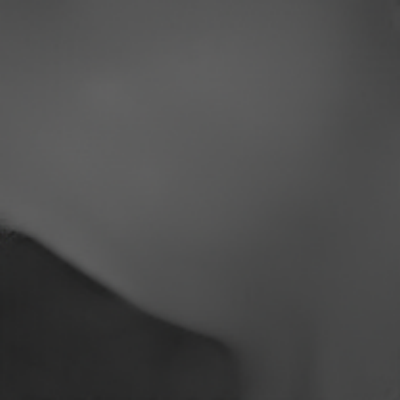}\\

\end{tabular}
\caption{The first image is the input $OC$ image. The input image is passed through $G_{VC}$ resulting in a synthetic $VC$, $VC_{syn}$. $OC_{rec}$ is  $VC_{syn}$ passed through  $G_{OC}$. Notice how this image does not have the same texture or the specular reflections as the input $OC$ image. Rather only the geometry between the two are the same. This geometry is reflected in $VC_{syn\_rec}$ which is obtained by passing $OC_{rec}$ through  $G_{VC}$.}
\label{fig:quad}
\end{figure}

\begin{figure*}[t!]
\begin{center}
\footnotesize
\includegraphics[width=0.85\textwidth]{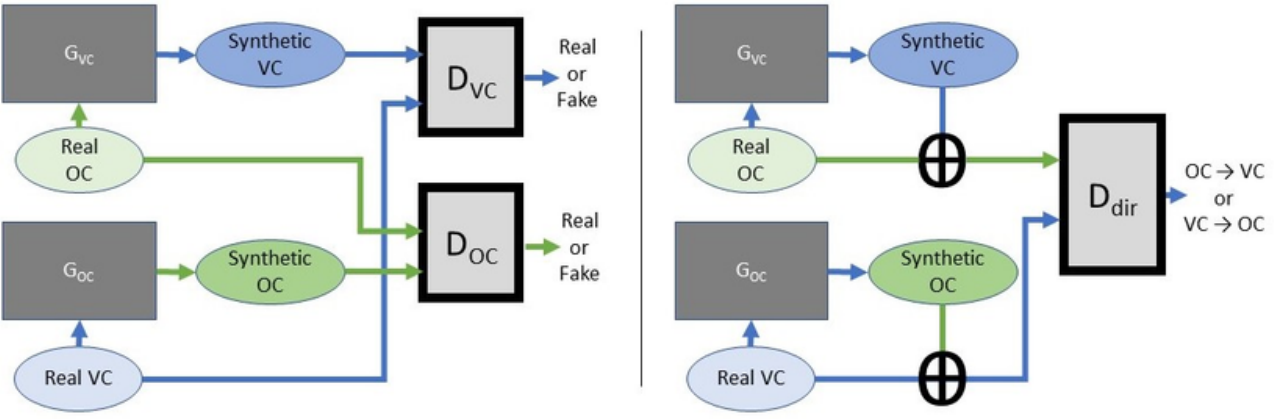}\\
\caption{The left image shows adversarial portion of CycleGAN to handle two GANs. Each generator acts independently without the cycle consistency losses included. The right image displays the architecture layout with a Directional Discriminator. Real OC and synthetic VC are concatenated and passed into the Directional Discriminator along with the concatenation of synthetic OC and real VC, creating a stronger connection between the two. This allows the Directional Discriminator to work with the paired information. In both cases, the discriminators only take into consideration the real distribution of real OC and VC along with the synthetic distributions produced by the generator from real OC and VC. Reconstructed images are not taken into account by the adversarial losses from these discriminators.}
\label{fig:directional}
\end{center}
\end{figure*}

\noindent
\textbf{Depth Reconstruction:}
Due to complexities in texture and lighting, traditional computer vision techniques do not work well for OC depth inference. Nadeem et al. \cite{nadeem2016computer} have introduced a non-parametric dictionary learning approach to infer depth information for a given OC video frame using only a VC RGB-Depth dictionary. However, due to the non-realistic rendering of depth cues in the VC RGB images, the inferred depth was inaccurate. Mahmood et al. \cite{mahmood2018unsupervised} have overcome this limitation by incorporating realistic depth cues, using inverse intensity fall-off in the rendered images. They created a transformer network that is trained on synthetic colon images. Given OC images, a GAN is used to first transform these images into a synthetic-like environment, which are then used to generate depth maps using a separate deep learning network. While this approach does a good job in removing patient-specific textures without requiring paired image data, it has difficulties removing specular reflections from the OC images. In addition, the resulting depth maps are not smooth and scale-consistent. 

Rau et al. \cite{rau2019implicit} have introduced a variant of pix2pix called extended pix2pix to produce OC depth maps. The extended pix2pix is a variant of the pix2pix model applied to colonoscopy depth reconstruction. Since a phantom and VC data was used to create paired depth and colon images, the network struggled with real OC data. To alleviate this problem, an extension was introduced that included real OC images for the GAN loss. Due to a lack of ground truth the L1 loss in pix2pix is ignored for these OC inputs. This allows the network to partially train on real colon images while not needing the corresponding ground truth. Their method, however, assumes a complete endoluminal view (tube-like structure) and fails otherwise. Chen et al. \cite{chen2019slam} have also used a pix2pix network to produce depth maps from a phantom model. They trained on VC with various realistic renderings that did not include any complex textures or specular reflections found in OC. Still, they were able to produce scale-consistent depth maps for a phantom and a porcine colon video sequence.

A deep learning method based on visual odometry has been presented  by Ma et al. \cite{ma2019real} to infer scale-consistent depth maps from OC video sequences. These scale-consistent depth maps are then passed into a SLAM algorithm \cite{schops2019surfelmeshing} to 3D reconstruct a colon mesh for surface coverage computation. Like most other methods, however, they assume a cylindrical topology and only caters to the endoluminal view. Furthermore, their method cannot handle specular highlights, occlusion and large camera movements, and requires preprocessing to mask these aspects.

\section{Data}
The OC and VC data was acquired at Stony Brook Hospital for 10 patients who underwent VC followed by OC (for polyp removal). The OC data contained 10 videos from OC procedures. These do not provide ground truth as the shape of the colon is different between VC and OC. The images taken from the videos were cropped to the borders of the frames. A cleansed 3D triangular mesh colon model was extracted from the 10 abdominal CT patient scans using a pipeline similar to Nadeem et al. \cite{nadeem2016computer}. The virtual colon was then loaded into the Blender$^1${\let\thefootnote\relax\footnote{{\hspace{-3mm}$^1$ www.blender.org}}} graphics software, and centerline flythrough videos of size 256$\times$256 pixels were rendered with two light sources on the sides of the virtual camera in order to replicate the endoscope and its environment. To incorporate more realistic depth cues, the inverse square fall-off property for the virtual lights was enabled \cite{mahmood2018unsupervised}. When training the network, both VC and OC images were downsampled to a size of 256$\times$256 pixels for computational efficiency. In total, 10 OC and VC videos were used with 5 of these used for training and the remaining three for testing and two for validation purposes. We captured 300 images from each OC and VC video, resulting in 1500 for training, 900 for testing and 600 for validation. Figure \ref{fig:pipeline} shows our end-to-end pipeline.

\begin{figure*}[t!]
\begin{center}
\setlength{\tabcolsep}{6pt}
\begin{tabular}{ccccccc}
& & \footnotesize{Histogram-Equalized} & & \footnotesize{Histogram-Equalized} & & \footnotesize{Histogram-Equalized} \\
OC Input & CycleGAN &  CycleGAN &XCycleGAN&XCycleGAN &XDCycleGAN & XDCycleGAN\\ \includegraphics[width=0.10\textwidth]{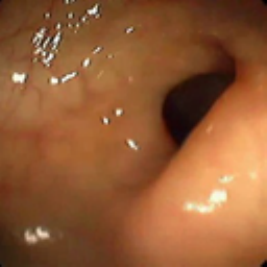}&
\includegraphics[width=0.10\textwidth]{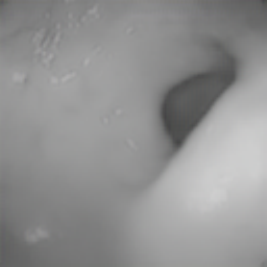}&
\includegraphics[width=0.10\textwidth]{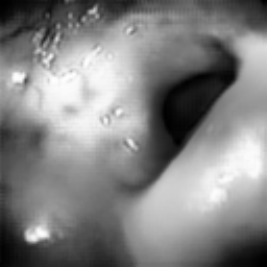}&
\includegraphics[width=0.10\textwidth]{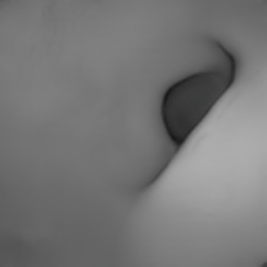}&
\includegraphics[width=0.10\textwidth]{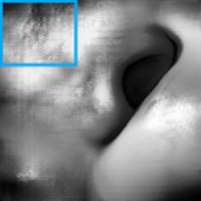}&
\includegraphics[width=0.10\textwidth]{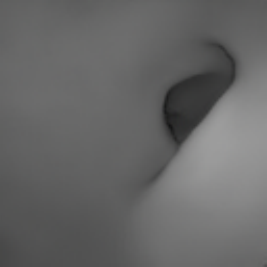}&
\includegraphics[width=0.10\textwidth]{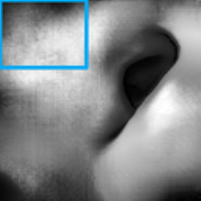}\\

\includegraphics[width=0.10\textwidth]{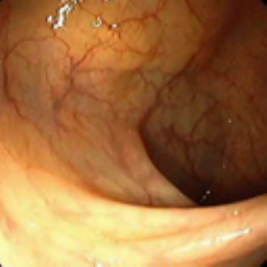}&
\includegraphics[width=0.10\textwidth]{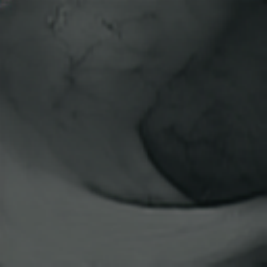}&
\includegraphics[width=0.10\textwidth]{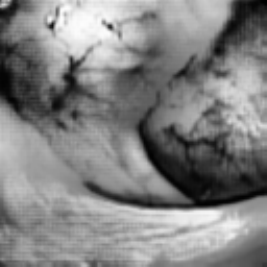}&
\includegraphics[width=0.10\textwidth]{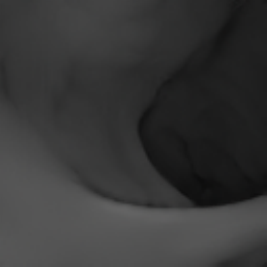}&
\includegraphics[width=0.10\textwidth]{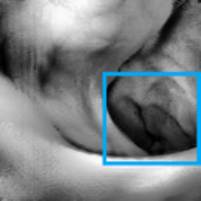}&
\includegraphics[width=0.10\textwidth]{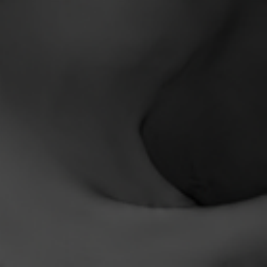}&
\includegraphics[width=0.10\textwidth]{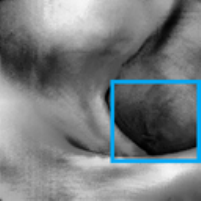}\\

\includegraphics[width=0.10\textwidth]{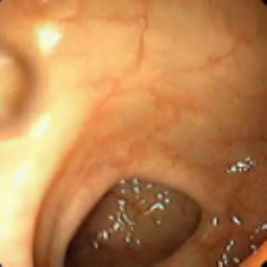}&
\includegraphics[width=0.10\textwidth]{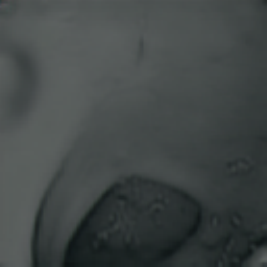}&
\includegraphics[width=0.10\textwidth]{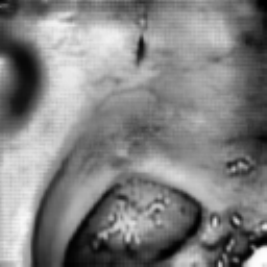}&
\includegraphics[width=0.10\textwidth]{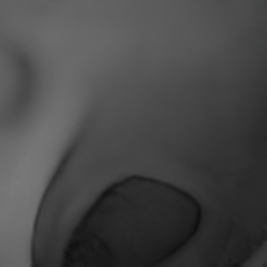}&
\includegraphics[width=0.10\textwidth]{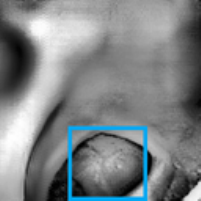}&
\includegraphics[width=0.10\textwidth]{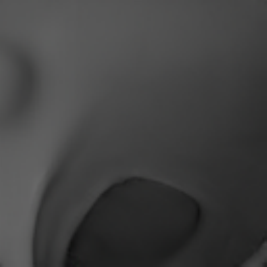}&
\includegraphics[width=0.10\textwidth]{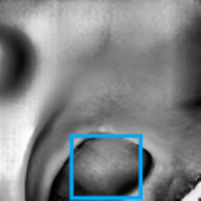}\\

\includegraphics[width=0.10\textwidth]{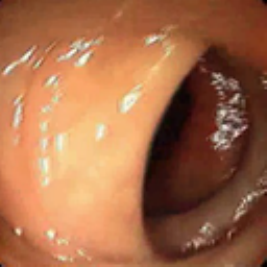}&
\includegraphics[width=0.10\textwidth]{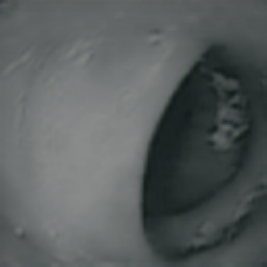}&
\includegraphics[width=0.10\textwidth]{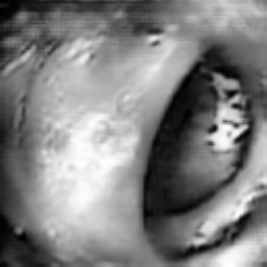}&
\includegraphics[width=0.10\textwidth]{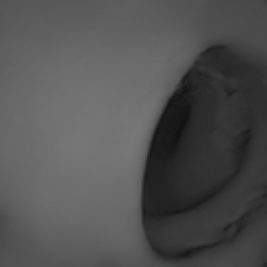}&
\includegraphics[width=0.10\textwidth]{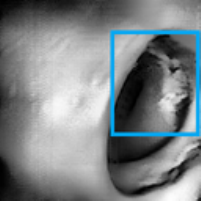}&

\includegraphics[width=0.10\textwidth]{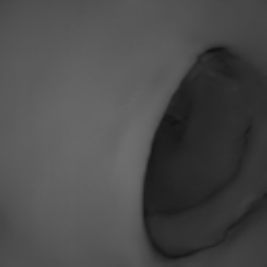}&
\includegraphics[width=0.10\textwidth]{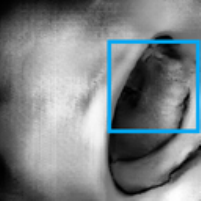}\\

\includegraphics[width=0.10\textwidth]{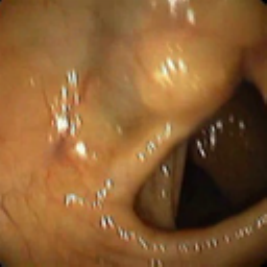}&
\includegraphics[width=0.10\textwidth]{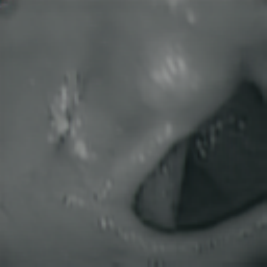}&
\includegraphics[width=0.10\textwidth]{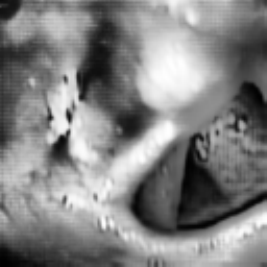}&
\includegraphics[width=0.10\textwidth]{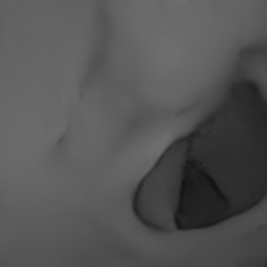}&
\includegraphics[width=0.10\textwidth]{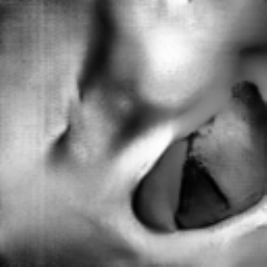}&

\includegraphics[width=0.10\textwidth]{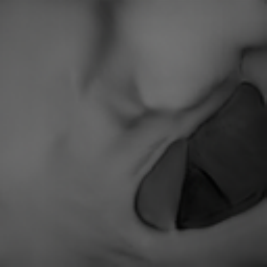}&
\includegraphics[width=0.10\textwidth]{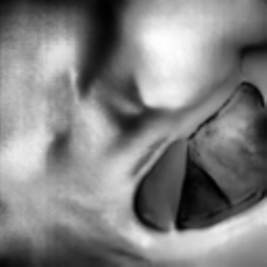}\\

\includegraphics[width=0.10\textwidth]{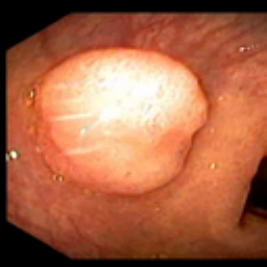}&
\includegraphics[width=0.10\textwidth]{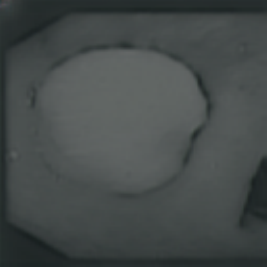}&
\includegraphics[width=0.10\textwidth]{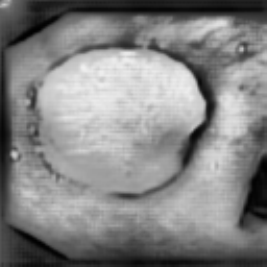}&
\includegraphics[width=0.10\textwidth]{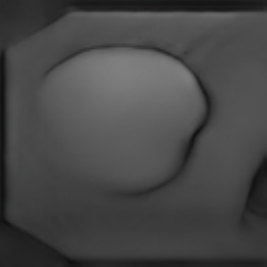}&
\includegraphics[width=0.10\textwidth]{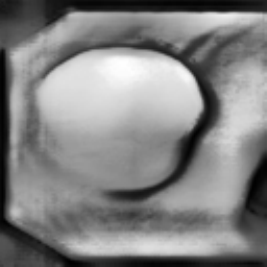}&
\includegraphics[width=0.10\textwidth]{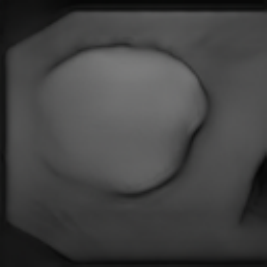}&
\includegraphics[width=0.10\textwidth]{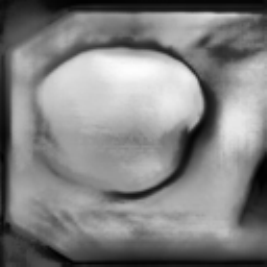}\\

\includegraphics[width=0.10\textwidth]{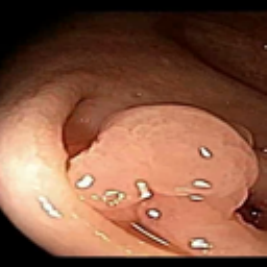}&
\includegraphics[width=0.10\textwidth]{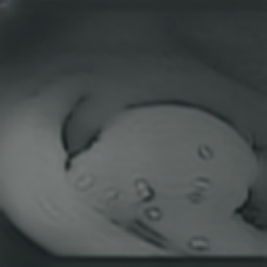}&
\includegraphics[width=0.10\textwidth]{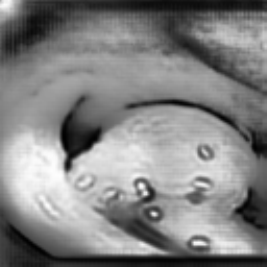}&
\includegraphics[width=0.10\textwidth]{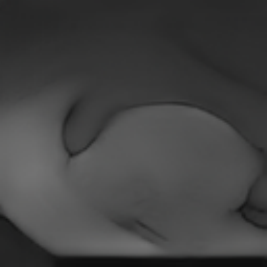}&
\includegraphics[width=0.10\textwidth]{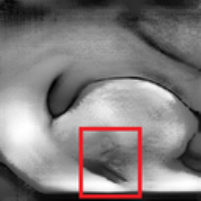}&
\includegraphics[width=0.10\textwidth]{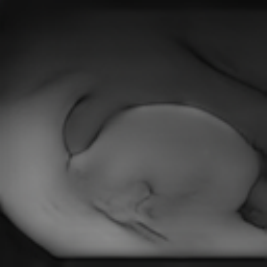}&
\includegraphics[width=0.10\textwidth]{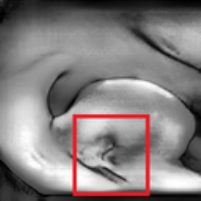}\\

\end{tabular}
\caption{Results from OC to VC. Blue boxes show areas where the XCycleGAN is unable to completely remove the specular reflections and texture, whereas the XDCycleGAN is able to remove these. The last two rows show polyps from \cite{mesejo2016computer} that XDCycleGAN can recreate polyps in VC.}
\label{fig:OC->VC}
\vspace{-2em}
\end{center}
\end{figure*}

\section{Method}

The CycleGAN network \cite{zhu2017unpaired} consists of two GANs with additional losses to combine the GANs into one model. We define $G$ as a generator, $G_{oc}$ as the generator from the GAN that produces OC images, and $G_{vc}$ as the generator that produces VC images. $D$, $D_{oc}$, and $D_{vc}$ represents discriminators for their corresponding generators. Similar to Zhu et al. \cite{zhu2017unpaired}, we represent the data distribution of domain $A$ as $y \backsim p(A)$ and the distribution of domain $B$ as $x \backsim p(B)$. The adversarial loss that is applied in GANs is as follows:
\begin{equation}
\begin{split}
     \mathcal{L}_{GAN}(G,D,A,B) = \  & \mathds{E}_{y \backsim p(A)} \big[ \textrm{log} (D(y))\big] + \\
    & \mathds{E}_{x \backsim p(B)} \big[\textrm{log}(1 - D(G(x))\big]
\end{split}
\end{equation}

The cycle consistency loss in CycleGANs links the two GANs to handle the image-to-image domain translation task. The cycle consistency loss is as follows:
\begin{equation}
    \mathcal{L}_{cyc}(G_a,G_b,A) = \mathds{E}_{y \backsim p(A)} \|y - G_a(G_b(y))\|_1
\end{equation}
where $\|\cdot\|_1$ is the $\ell 1$ norm, and $x \in a $. This loss is depicted on the left in Figure \ref{fig:difference}. The cycle consistency  loss is used for translating in both directions (i.e., A to B and B to A). The lossy transformations as seen in Figure \ref{fig:CLAHE} are not handled by the CycleGAN (as is previously shown \cite{chu2017cyclegan,porav2019reducing}). The cycle consistency loss requires OC images to be reconstructed from synthetic VC, $G_{vc}(OC)$. In order to handle this task, the network requires synthetic VC to store color, texture, and specular reflections so the synthetic VC can reconstruct the OC. To address this problem, we introduce the extended cycle consistency loss to help the network perform lossy translations. Still, there are textures and reflections that are passed into the VC domain and hence, a stronger link between OC and VC is required which is established via our Directional Discriminator that pairs OC and VC images.

\subsection{Extended Cycle Consistency Loss}
To address the OC features being embedded in VC, we propose a new loss to replace the cycle consistency loss in the OC domain, which we call the extended cycle consistency loss (Figure \ref{fig:difference}):

\footnotesize
\begin{equation}
    \mathcal{L}_{ excyc}(G_a,G_b,A) = \mathds{E}_{y \backsim p(A)} \|G_b(y) - G_b(G_a(G_b(y)))\|_1
\end{equation}
\normalsize

This loss has synthetic VC, $G_{vc}(OC)$, compared with reconstructed synthetic VC, $G_{vc}(G_{oc}(G_{vc}(OC)))$. In other words, the extended cycle consistency loss is enforcing the structure captured in the VC domain to be the same between OC and the reconstructed OC, $G_{oc}(G_{vc}(OC))$. This loss is depicted pictorially on the right in Figure \ref{fig:difference}. Figure \ref{fig:quad} shows how the extended cycle consistency loss allows the reconstructed OC to have different textures and lighting than the original OC input. When applying this loss to the CycleGAN, we call it the extended CycleGAN (XCycleGAN).

The network, the way it is, has the reconstructed OC, $G_{oc}(G_{vc}(OC))$, unrestrained. Since this image is supposed to look like an OC image, an additional OC discriminator is added and a GAN loss is applied. In addition, Zhu et al. \cite{zhu2017unpaired}  have mentioned the use of an identity loss that compares OC and $G_{OC}(OC)$ to retain color when reconstructing. This loss is removed as we do not wish to retain color information for OC but is kept on the VC side to retain the color there. Thus, $\mathcal{L}_{iden}(A)= \mathds{E}_{y \backsim p(A)} |G_{a}(y)-y|$, is a loss included for the XCycleGAN.

\begin{figure}[t!]
\begin{center}
\setlength{\tabcolsep}{1pt}
\begin{tabular}{ccc}

VC Input & ~~~~~CycleGAN & ~~~~~XDCycleGAN \\ 
\includegraphics[width=0.1\textwidth]{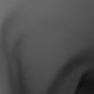}& ~~~~~
\includegraphics[width=0.1\textwidth]{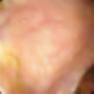}& ~~~~~
\includegraphics[width=0.1\textwidth]{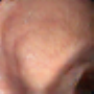}\\
\includegraphics[width=0.1\textwidth]{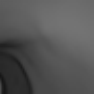}&~~~~~
\includegraphics[width=0.1\textwidth]{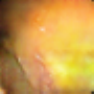}&~~~~~
\includegraphics[width=0.1\textwidth]{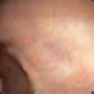}\\
\includegraphics[width=0.1\textwidth]{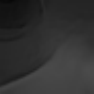}&~~~~~
\includegraphics[width=0.1\textwidth]{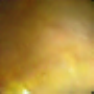}&~~~~~
\includegraphics[width=0.1\textwidth]{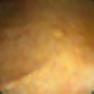}\\
\includegraphics[width=0.1\textwidth]{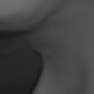}&~~~~~
\includegraphics[width=0.1\textwidth]{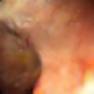}&~~~~~
\includegraphics[width=0.1\textwidth]{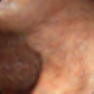}\\
\end{tabular}

\begin{tabular}{cccc}
VC Input & OC Output & VC Input & OC Output\\ 
\includegraphics[width=0.1\textwidth]{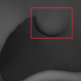}& \includegraphics[width=0.1\textwidth]{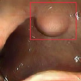}&
\includegraphics[width=0.1\textwidth]{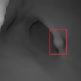}& \includegraphics[width=0.1\textwidth]{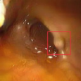} \\
\includegraphics[width=0.1\textwidth]{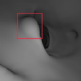}& \includegraphics[width=0.1\textwidth]{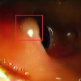}&
\includegraphics[width=0.1\textwidth]{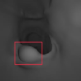}& \includegraphics[width=0.1\textwidth]{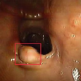} \\
\includegraphics[width=0.1\textwidth]{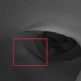}& \includegraphics[width=0.1\textwidth]{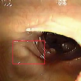}&
\includegraphics[width=0.1\textwidth]{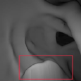}& \includegraphics[width=0.1\textwidth]{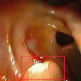} \\
\end{tabular}
\caption{Results from VC to OC image translation. The synthetic OC images from the CycleGAN turn structures into texture where the XDCycleGAN does not do this. Below that, we display polyps created in VC with augmneted textures, colors, and specular reflections by our XDCycleGAN.}
\vspace{-1em}

\label{fig:vc2oc}
\end{center}
\end{figure}

\begin{figure*}[t!]
\begin{center}
\setlength{\tabcolsep}{4pt}
\begin{tabular}{cccccccccccc}

\rotatebox{90}{\rlap{\small ~~~~~~~~OC}}&

\includegraphics[width=0.075\textwidth]{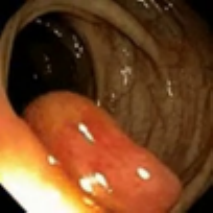}&
\includegraphics[width=0.075\textwidth]{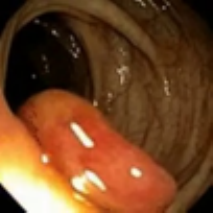}&
\includegraphics[width=0.075\textwidth]{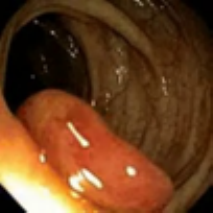}&
\includegraphics[width=0.075\textwidth]{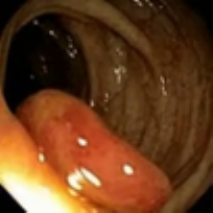}&
\ldots&
\includegraphics[width=0.075\textwidth]{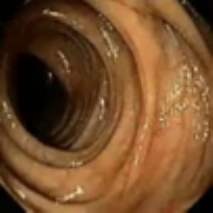}&
\includegraphics[width=0.075\textwidth]{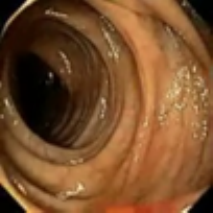}&
\includegraphics[width=0.075\textwidth]{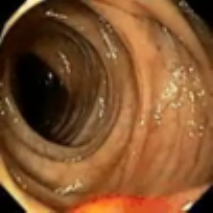}&
\includegraphics[width=0.075\textwidth]{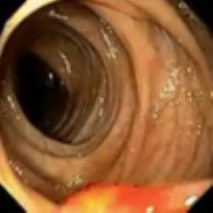}\\

\rotatebox{90}{\rlap{\scriptsize Mahmood \cite{mahmood2018unsupervised}}}&

\includegraphics[width=0.075\textwidth]{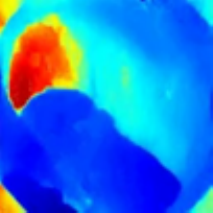}&
\includegraphics[width=0.075\textwidth]{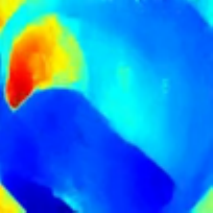}&
\includegraphics[width=0.075\textwidth]{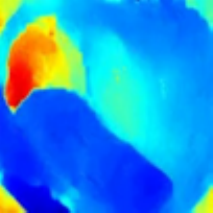}&
\includegraphics[width=0.075\textwidth]{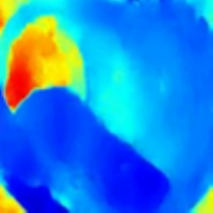}&
\ldots&
\includegraphics[width=0.075\textwidth]{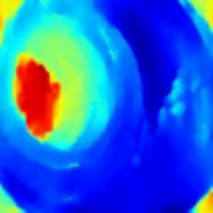}&
\includegraphics[width=0.075\textwidth]{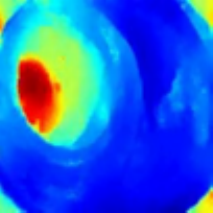}&
\includegraphics[width=0.075\textwidth]{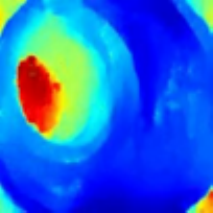}&
\includegraphics[width=0.075\textwidth]{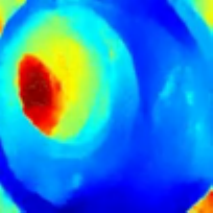}\\
\rotatebox{90}{\rlap{\small ~~~~Ours}}&

\includegraphics[width=0.075\textwidth]{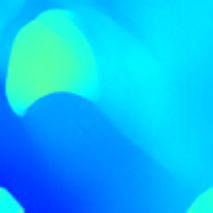}&
\includegraphics[width=0.075\textwidth]{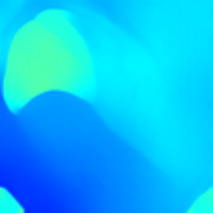}&
\includegraphics[width=0.075\textwidth]{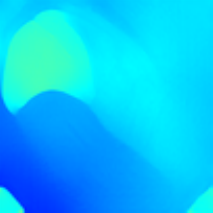}&
\includegraphics[width=0.075\textwidth]{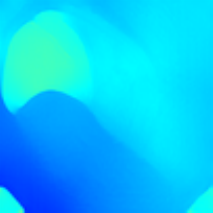}&
\ldots&
\includegraphics[width=0.075\textwidth]{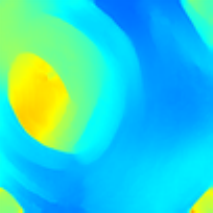}&
\includegraphics[width=0.075\textwidth]{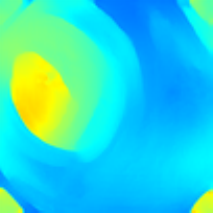}&
\includegraphics[width=0.075\textwidth]{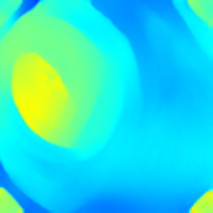}&
\includegraphics[width=0.075\textwidth]{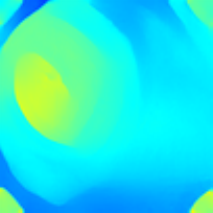}\\

\hline \\[-1.0em]

\rotatebox{90}{\rlap{\tiny ~~~ Textured VC}}&
\includegraphics[width=0.075\textwidth]{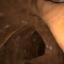}&
\includegraphics[width=0.075\textwidth]{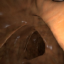}&
\includegraphics[width=0.075\textwidth]{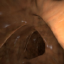}&
\includegraphics[width=0.075\textwidth]{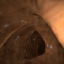}&
\ldots&
\includegraphics[width=0.075\textwidth]{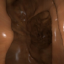}&
\includegraphics[width=0.075\textwidth]{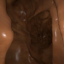}&
\includegraphics[width=0.075\textwidth]{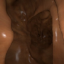}&
\includegraphics[width=0.075\textwidth]{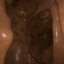}\\

\rotatebox{90}{\rlap{ \tiny ~ Ground Truth}}&
\includegraphics[width=0.075\textwidth]{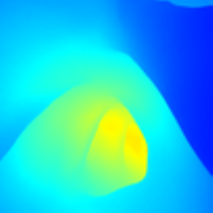}&
\includegraphics[width=0.075\textwidth]{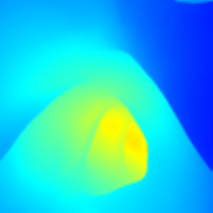}&
\includegraphics[width=0.075\textwidth]{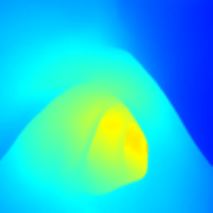}&
\includegraphics[width=0.075\textwidth]{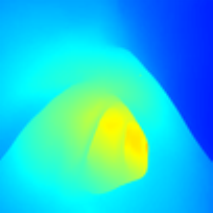}&
\ldots&
\includegraphics[width=0.075\textwidth]{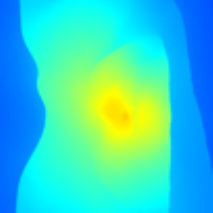}&
\includegraphics[width=0.075\textwidth]{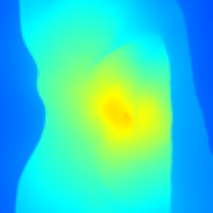}&
\includegraphics[width=0.075\textwidth]{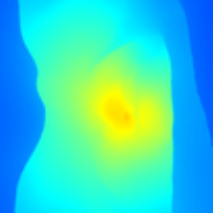}&
\includegraphics[width=0.075\textwidth]{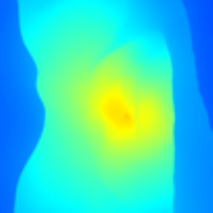}\\

\rotatebox{90}{\rlap{\footnotesize ~~~Ma \cite{ma2019real}}}&
\includegraphics[width=0.075\textwidth]{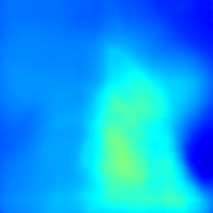}&
\includegraphics[width=0.075\textwidth]{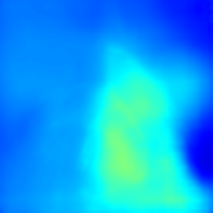}&
\includegraphics[width=0.075\textwidth]{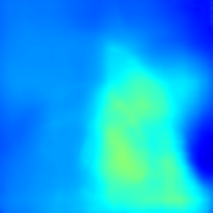}&
\includegraphics[width=0.075\textwidth]{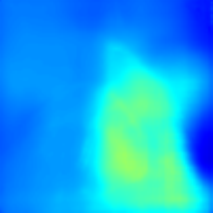}&
\ldots&
\includegraphics[width=0.075\textwidth]{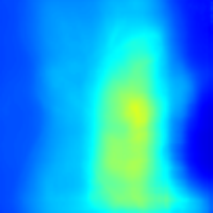}&
\includegraphics[width=0.075\textwidth]{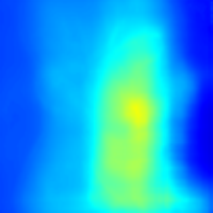}&
\includegraphics[width=0.075\textwidth]{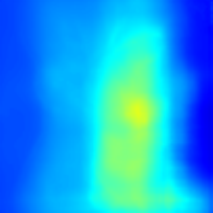}&
\includegraphics[width=0.075\textwidth]{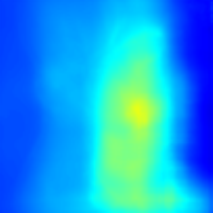}\\

\rotatebox{90}{\rlap{\small ~~~~Ours}}&
\includegraphics[width=0.075\textwidth]{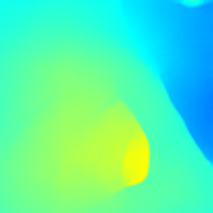}&
\includegraphics[width=0.075\textwidth]{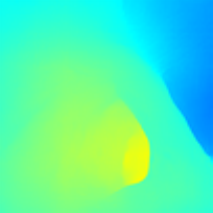}&
\includegraphics[width=0.075\textwidth]{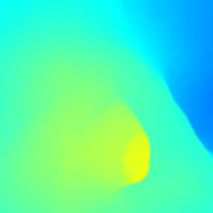}&
\includegraphics[width=0.075\textwidth]{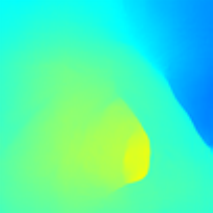}&
\ldots&
\includegraphics[width=0.075\textwidth]{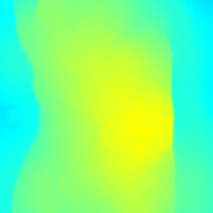}&
\includegraphics[width=0.075\textwidth]{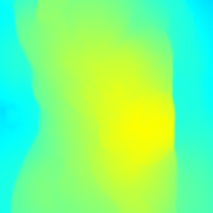}&
\includegraphics[width=0.075\textwidth]{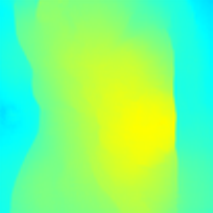}&
\includegraphics[width=0.075\textwidth]{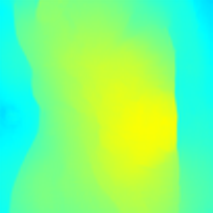}\\

\hline \\[-1.0em]

\rotatebox{90}{\rlap{\small ~~~~~OC}}&

\includegraphics[width=0.075\textwidth]{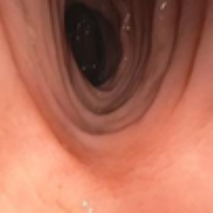}&
\includegraphics[width=0.075\textwidth]{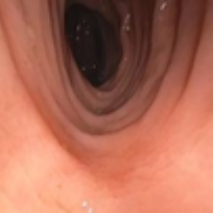}&
\includegraphics[width=0.075\textwidth]{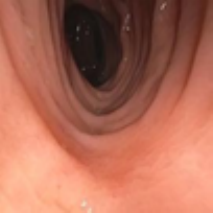}&
\includegraphics[width=0.075\textwidth]{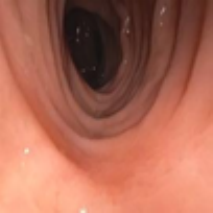}&
\ldots&
\includegraphics[width=0.075\textwidth]{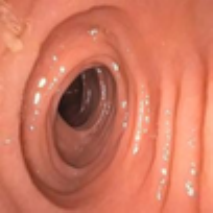}&
\includegraphics[width=0.075\textwidth]{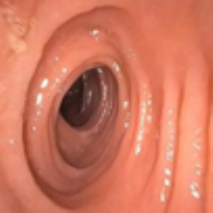}&
\includegraphics[width=0.075\textwidth]{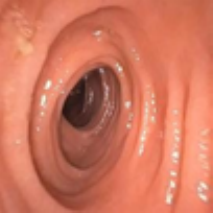}&
\includegraphics[width=0.075\textwidth]{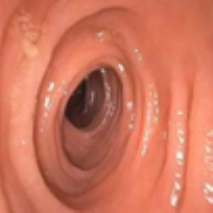}\\

\rotatebox{90}{\rlap{\small ~~Ma \cite{ma2019real}}}&
\includegraphics[width=0.075\textwidth]{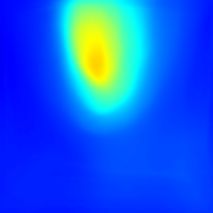}&
\includegraphics[width=0.075\textwidth]{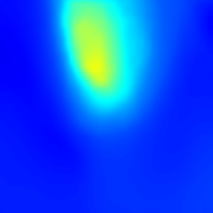}&
\includegraphics[width=0.075\textwidth]{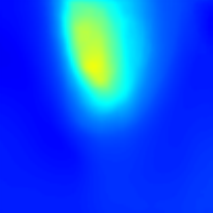}&
\includegraphics[width=0.075\textwidth]{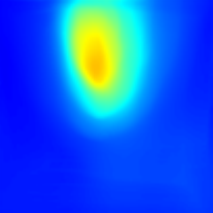}&
\ldots&
\includegraphics[width=0.075\textwidth]{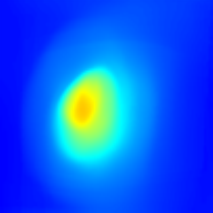}&
\includegraphics[width=0.075\textwidth]{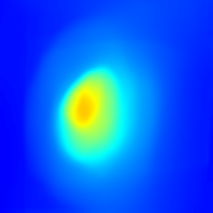}&
\includegraphics[width=0.075\textwidth]{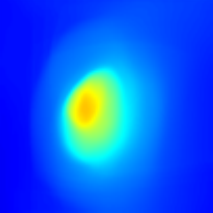}&
\includegraphics[width=0.075\textwidth]{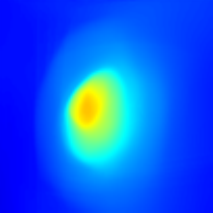}\\

\rotatebox{90}{\rlap{\small ~~~~Ours}}&
\includegraphics[width=0.075\textwidth]{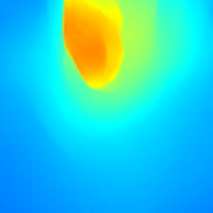}&
\includegraphics[width=0.075\textwidth]{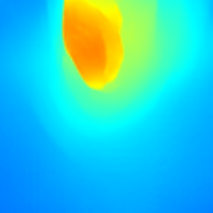}&
\includegraphics[width=0.075\textwidth]{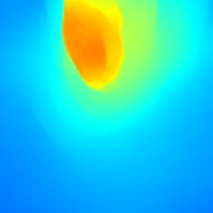}&
\includegraphics[width=0.075\textwidth]{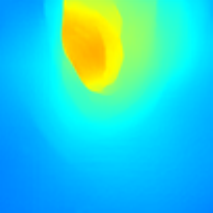}&
\ldots&
\includegraphics[width=0.075\textwidth]{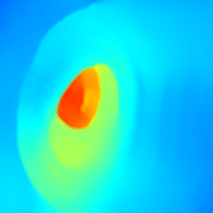}&
\includegraphics[width=0.075\textwidth]{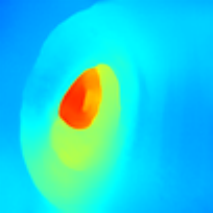}&
\includegraphics[width=0.075\textwidth]{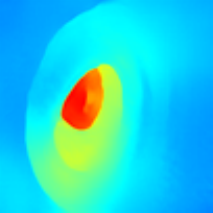}&
\includegraphics[width=0.075\textwidth]{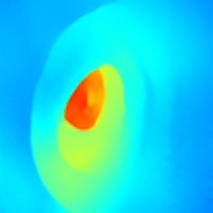}\\

\hline \\[-1.0em]

\rotatebox{90}{\rlap{\scriptsize ~~Phantom}}&
\includegraphics[width=0.075\textwidth]{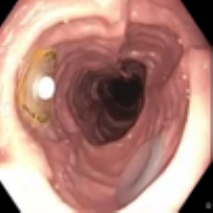}&
\includegraphics[width=0.075\textwidth]{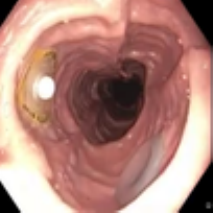}&
\includegraphics[width=0.075\textwidth]{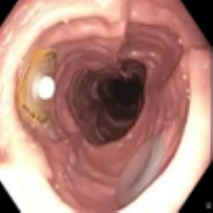}&
\includegraphics[width=0.075\textwidth]{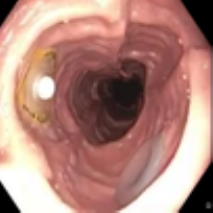}&
\ldots&
\includegraphics[width=0.075\textwidth]{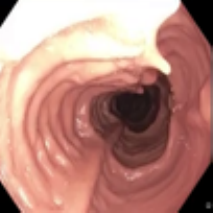}&
\includegraphics[width=0.075\textwidth]{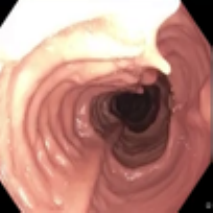}&
\includegraphics[width=0.075\textwidth]{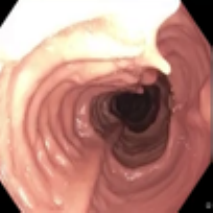}&
\includegraphics[width=0.075\textwidth]{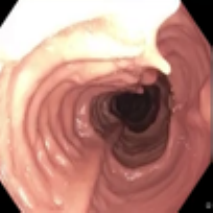}\\

\rotatebox{90}{\rlap{\footnotesize ~Chen \cite{chen2019slam}}}&
\includegraphics[width=0.075\textwidth]{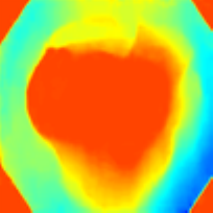}&
\includegraphics[width=0.075\textwidth]{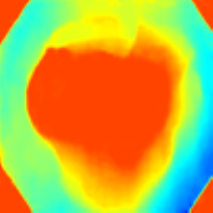}&
\includegraphics[width=0.075\textwidth]{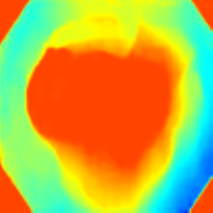}&
\includegraphics[width=0.075\textwidth]{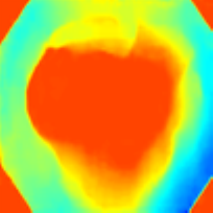}&
\ldots&
\includegraphics[width=0.075\textwidth]{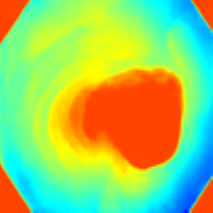}&
\includegraphics[width=0.075\textwidth]{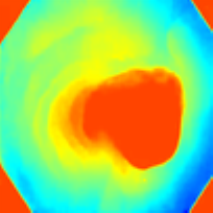}&
\includegraphics[width=0.075\textwidth]{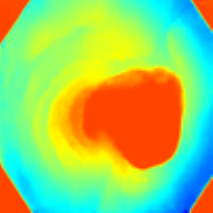}&
\includegraphics[width=0.075\textwidth]{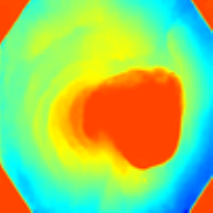}\\

\rotatebox{90}{\rlap{\footnotesize ~~~~Ours}}&
\includegraphics[width=0.075\textwidth]{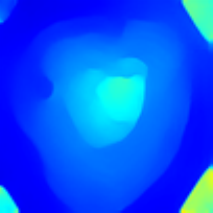}&
\includegraphics[width=0.075\textwidth]{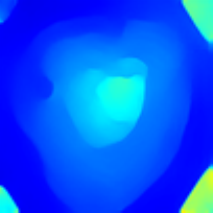}&
\includegraphics[width=0.075\textwidth]{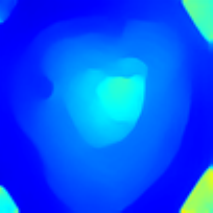}&
\includegraphics[width=0.075\textwidth]{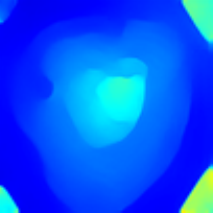}&
\ldots&
\includegraphics[width=0.075\textwidth]{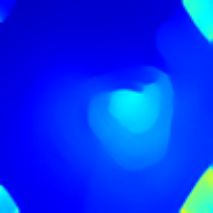}&
\includegraphics[width=0.075\textwidth]{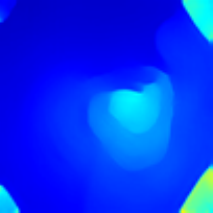}&
\includegraphics[width=0.075\textwidth]{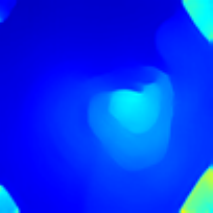}&
\includegraphics[width=0.075\textwidth]{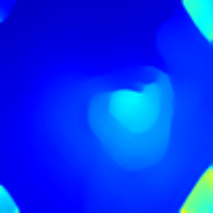}\\

\end{tabular}
\caption{Scale-consistent depth inference on video sequences. The first (top) dataset contains a polyp video sequence with Mahmood et al.'s \cite{mahmood2018unsupervised} and our results. The second dataset is successive frames from our textured VC video flythrough and the corresponding results from Ma et al. \cite{ma2019real} and ours. The SSIM for Ma et al.'s approach is 0.637, whereas ours is 0.918. Ma et al. assume as input a chunk of successive video frames with cylindrical topology (endoluminal) view with the specular reflections and occlusions masked out.  The third dataset show another sequence assuming Ma et al.'s input \cite{ma2019real} and our corresponding results. The final dataset, shows Chen et al.'s phantom model \cite{chen2019slam} along with their results and our results. Complete videos and additional sequences are shown in the supplementary material. }
\label{fig:time}
\end{center}
\end{figure*}

\subsection{Directional Discriminator}
To create a stronger link between OC and VC, our approach uses a single Directional Discriminator rather than two as shown in Figure \ref{fig:directional}. Since only the number of input channels of the discriminator is changed, the Directional Discriminator reduces the memory needed for the network. Our Directional Discriminator only required 17.4 MB, whereas a single CycleGAN discriminator required 11.1 MB (altogether 22.2MB).

GANs are based around the idea of creating a two player adversarial game between the generator and the discriminator. In our model, we wish to create a stronger relationship between the image generators by creating a three player game. The players of this game are two generators ($G_{a}$,$G_{b}$) and a discriminator ($D$). Similar to conditional GANs, $G_{a}$ will give its input and output to $D$ trying to convince $D$ that its output came from $G_{b}$. $G_{b}$ does the same task except it makes its input-output pair resemble $G_{a}$'s. Since, $G_a$'s input domain is $G_B$'s output domain and $G_a$'s output domain is $G_b$'s input domain, the discriminator ends up discerning which generator is used since the input domains are fixed as shown on the right in Figure \ref{fig:directional}. By trying to differentiate the generators, the discriminator is essentially differentiating the direction of the translation. For example, if we look at OC and VC image translation, the discriminator would be differentiating the following pairs \{OC, synthetic VC\} and \{synthetic OC, VC\}. Thus, the discriminator ends up discerning the direction of the translation. When this model converges, the synthesized images will need to reflect the real distribution of their corresponding domains, while also giving the discriminator paired information to work with. This creates a stronger connection between the two generators, while eliminating the need for two discriminators. The loss for this Directional Discriminator is:
\begin{equation}
\begin{split}
    \mathcal{L}_{dir}(G_a,G_b,D,A,B) = \  & \mathds{E}_{y \backsim p(A)} \big[ \textrm{log} (D(y,G_b(y))\big] + \\
    & \mathds{E}_{x \backsim p(B)} \big[1- \textrm{log} (D(G_a(x),x)\big]
\end{split}
\end{equation}

The combination of our Directional Discriminator and extended cycle consistency loss produces the Extended and Directional CycleGAN (XDCycleGAN). The total objective loss function for XDCycleGAN is:

\begin{equation}
\begin{split}
    \mathcal{L} = & \lambda \mathcal{L}_{excyc}(G_{oc},G_{vc},I_{oc}) +  \lambda \mathcal{L}_{cyc}(G_{vc},G_{oc},I_{vc})  \\
    & + \mathcal{L}_{dir}(G_{oc},G_{vc},D_{dir},I_{oc},I_{vc}) \\
    & + \mathcal{L}_{dir}(G_{vc},G_{oc},D_{dir},I_{vc},I_{oc}) \\
    &  + \alpha \mathcal{L}_{GAN}(G_{oc},D_{oc},I_{oc},G_{oc}(G_{vc}(I_{oc})))\\
         &  + \gamma \mathcal{L}_{iden}(VC),
\end{split}
\end{equation}
where $\alpha,\lambda,$ and $\gamma$ are constant weights. For both OC to VC rendering and OC to scale consistent depth maps, we train the network for 200 epochs with $\alpha =0.5$, $\lambda = 10$, and $\gamma = 5$. We add spectral normalization \cite{miyato2018spectral} to each layer of the discriminators for better network stability.

\section{Experimental Results}
To clearly emphasize the texture and specular highlights in OC to VC translation, we show our results in Figure \ref{fig:OC->VC} by training the CycleGAN, XCycleGAN, and XDCycleGAN on OC and rendered VC data; in depth maps the texture and highlights are slightly difficult to visualize. To further highlight the embedding of the textures and lighting, histogram equalization is applied to the output VC images. For all the OC images, it is clearly seen that the histogram-equalized CycleGAN images embed the specular reflection and textures. In most cases these artifacts are visible in the VC images. We further point out that there are textures and lighting seen in the histogram-equalized XCycleGAN which are retained from the input, which the XDCycleGAN is able to remove. These cases are marked by the blue boxes. The last two rows in Figure \ref{fig:OC->VC} show how our network is able to recreate the polyps in VC. In the third last row, the XDCycleGAN shows its superior understanding of the geometry from OC. It recovers the shape of the polyp better than the XCycleGAN, which makes it appear much flatter. In the last row the more intricate geometry of the polyp is captured by the XDCycleGAN and is highlighted with a red box.

In the VC to OC image translation, our network also does better than  CycleGAN, as seen in Figure \ref{fig:vc2oc}. We observe that  CycleGAN takes structures in the VC domain and turns these into texture, whereas the XDCycleGAN retains all of the structure in the VC domain. In order to demonstrate the benefits of the VC to OC translation, we generate polyps by adding bumps with specific shapes and endoscope orientations in VC and augment them with textures and specular reflection also shown in Figure \ref{fig:vc2oc}. The cycle consistency accuracy for depth in the VC $\rightarrow$ OC $\rightarrow$ VC case for the CycleGAN is 7.74$\pm$6.07, for the XCycleGAN is 6.84$\pm$5.39, and for the XDCycleGAN is 6.34$\pm$3.73.

\begin{figure}[t!]
\begin{center}
\setlength{\tabcolsep}{1pt}
\begin{tabular}{cccc}
OC Input &  XDCycleGAN & OC Input &  XDCycleGAN\\ 
\includegraphics[width=0.10\textwidth]{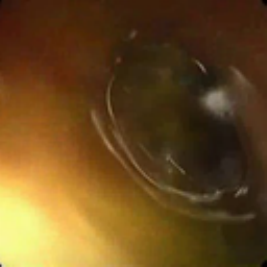}&
\includegraphics[width=0.10\textwidth]{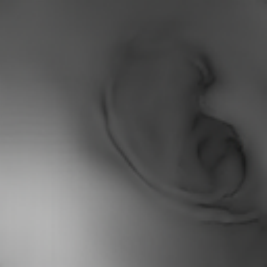}&
\includegraphics[width=0.10\textwidth]{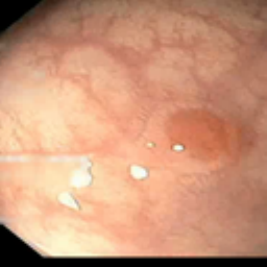}&
\includegraphics[width=0.10\textwidth]{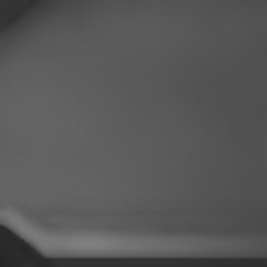}\\
\includegraphics[width=0.10\textwidth]{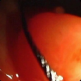}&
\includegraphics[width=0.10\textwidth]{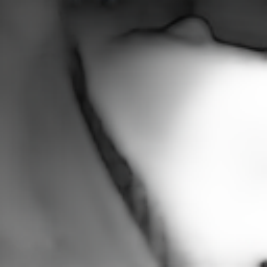}&
\includegraphics[width=0.10\textwidth]{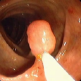}&
\includegraphics[width=0.10\textwidth]{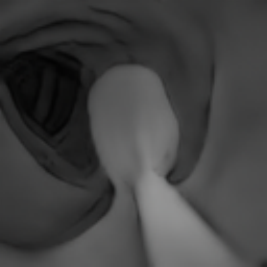}\\

\end{tabular}
\caption{Failure cases of the XDCycleGAN. These include frames with fluid movement, motion blur or occlusion, as well as instruments. The top-right image (from \cite{mesejo2016computer}) shows how our network can mistake polyps for texture if it is surrounded by texture.}
\vspace{-1em}
\label{fig:fail}
\end{center}
\end{figure}

\noindent
\textbf{Scale-Consistent Depth Inference}
We train our XDCycleGAN on VC depth maps to demonstrate its ability to infer scale-consistent depth information. This depth inference is a first step for 3D reconstruction, and once these depth maps are attained, standard SLAM algorithms \cite{schops2019surfelmeshing,whelan2015elasticfusion} can be used. Figure \ref{fig:time} shows various time sequences and our results on that input. In our first test, we compare with Mahmood et al. \cite{mahmood2018unsupervised} and show that our depth maps are much smoother. Their method shows issues by treating some of the specular reflections as a change in depth and failing in scale consistency. The XDCycleGAN, on the other hand, ignores these reflections and produces smoother depth maps.

For our second test, we use 2000 video frames produced from a manually textured virtual colon from VC. The colon is textured by blending various snapshots of textures found in OC, and ground truth can be attained.  For quantitative analysis, we analyze the average SSIM score across the 2000 frames to test structural similarity. Our approach got an average SSIM score of 0.918, which indicates that our network is able to capture the structure of these frames rather well. We also ran Ma et al.'s approach \cite{ma2019real} on our ground truth. The SSIM score was 0.637. Their score is low as the depth maps are smoothed out and most of the critical geometry/structures are lost. They also assume the endoluminal view which, if invalidated, propagates the error to the following frames. In addition to SSIM, RMSE for the depth maps with our approach is 31.25$\pm$6.76 as compared to 92.67$\pm$10.32 for Ma et al. \cite{ma2019real}.

To be fair, we also compared Ma et al.'s approach \cite{ma2019real} on OC video sequences where their input assumptions hold. The depth maps produced from their pipeline does not recover the geometry of most haustral folds as accurately as XDCycleGAN. Finally, we compare with Chen et al.'s method \cite{chen2019slam} on their phantom colon and show that our model is robust, and is comparable even on data that we did not cater for during training.

\section{Limitations and Future work}
As with most deep learning frameworks there are failure cases with this network, as shown in Figure \ref{fig:fail}. When there is heavy occlusion, blurring, or fluid motion the network tries to reconstruct the image regardless and infers random geometry. To address these artifacts, in the future we will incorporate an image quality control framework to detect and remove these frames. Moreover, our pipeline does not cater to the instruments in the OC images since our VC counterpart does not have equivalent representation. Adding instrument models in the VC domain may be one way to recognize instruments in the OC images. Furthermore, if a polyp strongly blends in with the textures of the colon wall and its protrusion from the wall is hard to notice, the XDCycleGAN removes both the polyp and the texture. The polyp, however, may be visible in the earlier frames and a temporal component may resolve this. Finally, we will undertake a more through analysis of polyp detection and segmentation pipelines with our polyp data augmentation/synthesis.

\section*{Acknowledgement}
We would like to thank Dr. Sarah K. McGill (UNC Chapel Hill) for granting access to the OC videos and to Ruibin Ma for comparisons. This research was funded by NSF grants NRT1633299, CNS1650499, OAC1919752, and ICER1940302, the NIH/NCI Cancer Center Support Grant P30 CA008748, NIH/NHLBI under Award U01HL127522, the Center for Biotechnology – NY State Center for Advanced Technology, Stony Brook University, Cold Spring Harbor Lab, Brookhaven National Lab, Feinstein Institute for Medical Research, and NY State Department of Economic Development under Contract C14051. The content is solely the responsibility of the authors and does not necessarily represent the official views of the NIH.


\begin{thebibliography}{10}\itemsep=-1pt
\small 
\bibitem{bray2018global}
Freddie Bray, Jacques Ferlay, Isabelle Soerjomataram, Rebecca~L Siegel,
  Lindsey~A Torre, and Ahmedin Jemal.
\newblock Global cancer statistics 2018: {GLOBOCAN} estimates of incidence and
  mortality worldwide for 36 cancers in 185 countries.
\newblock {\em CA: A Cancer Journal for Clinicians}, 68(6):394--424, 2018.

\bibitem{chen2019slam}
Richard~J Chen, Taylor~L Bobrow, Thomas Athey, Faisal Mahmood, and Nicholas~J
  Durr.
\newblock Slam endoscopy enhanced by adversarial depth prediction.
\newblock {\em arXiv preprint arXiv:1907.00283}, 2019.

\bibitem{chu2017cyclegan}
Casey Chu, Andrey Zhmoginov, and Mark Sandler.
\newblock Cyclegan, a master of steganography.
\newblock {\em arXiv preprint arXiv:1712.02950}, 2017.

\bibitem{donahue2016adversarial}
Jeff Donahue, Philipp Kr{\"a}henb{\"u}hl, and Trevor Darrell.
\newblock Adversarial feature learning.
\newblock {\em International Conference on Learning Representations (ICLR)},
  2017.

\bibitem{dumoulin2016adversarially}
Vincent Dumoulin, Ishmael Belghazi, Ben Poole, Olivier Mastropietro, Alex Lamb,
  Martin Arjovsky, and Aaron Courville.
\newblock Adversarially learned inference.
\newblock {\em International Conference on Learning Representations (ICLR)},
  2017.

\bibitem{goodfellow2014generative}
Ian Goodfellow, Jean Pouget-Abadie, Mehdi Mirza, Bing Xu, David Warde-Farley,
  Sherjil Ozair, Aaron Courville, and Yoshua Bengio.
\newblock Generative adversarial nets.
\newblock {\em Advances in neural information processing systems}, pages
  2672--2680, 2014.

\bibitem{isola2017image}
Phillip Isola, Jun-Yan Zhu, Tinghui Zhou, and Alexei~A Efros.
\newblock Image-to-image translation with conditional adversarial networks.
\newblock {\em Proceedings of the IEEE Conference on Computer Vision and
  Pattern Recognition}, pages 1125--1134, 2017.

\bibitem{kim2017learning}
Taeksoo Kim, Moonsu Cha, Hyunsoo Kim, Jung~Kwon Lee, and Jiwon Kim.
\newblock Learning to discover cross-domain relations with generative
  adversarial networks.
\newblock {\em Proceedings of the 34th International Conference on Machine
  Learning}, 70:1857--1865, 2017.

\bibitem{ledig2017photo}
Christian Ledig, Lucas Theis, Ferenc Husz{\'a}r, Jose Caballero, Andrew
  Cunningham, Alejandro Acosta, Andrew Aitken, Alykhan Tejani, Johannes Totz,
  Zehan Wang, et~al.
\newblock Photo-realistic single image super-resolution using a generative
  adversarial network.
\newblock {\em Proceedings of the IEEE Conference on Computer Vision and
  Pattern Recognition}, pages 4681--4690, 2017.

\bibitem{long2015fully}
Jonathan Long, Evan Shelhamer, and Trevor Darrell.
\newblock Fully convolutional networks for semantic segmentation.
\newblock {\em Proceedings of the IEEE Conference on Computer Vision and
  Pattern Recognition}, pages 3431--3440, 2015.

\bibitem{lu2018attribute}
Yongyi Lu, Yu-Wing Tai, and Chi-Keung Tang.
\newblock Attribute-guided face generation using conditional cyclegan.
\newblock {\em Proceedings of the European Conference on Computer Vision
  (ECCV)}, pages 282--297, 2018.

\bibitem{ma2019real}
Ruibin Ma, Rui Wang, Stephen Pizer, Julian Rosenman, Sarah~K McGill, and
  Jan-Michael Frahm.
\newblock Real-time 3d reconstruction of colonoscopic surfaces for determining
  missing regions.
\newblock {\em International Conference on Medical Image Computing and
  Computer-Assisted Intervention}, pages 573--582, 2019.

\bibitem{mahmood2018unsupervised}
Faisal Mahmood, Richard Chen, and Nicholas~J Durr.
\newblock Unsupervised reverse domain adaptation for synthetic medical images
  via adversarial training.
\newblock {\em IEEE Transactions on Medical Imaging}, 37(12):2572--2581, 2018.

\bibitem{mathieu2015deep}
Michael Mathieu, Camille Couprie, and Yann LeCun.
\newblock Deep multi-scale video prediction beyond mean square error.
\newblock {\em arXiv preprint arXiv:1511.05440}, 2015.

\bibitem{mesejo2016computer}
Pablo Mesejo, Daniel Pizarro, Armand Abergel, Olivier Rouquette, Sylvain
  Beorchia, Laurent Poincloux, and Adrien Bartoli.
\newblock Computer-aided classification of gastrointestinal lesions in regular
  colonoscopy.
\newblock {\em IEEE Transactions on Medical Imaging}, 35(9):2051--2063, 2016.

\bibitem{mirza2014conditional}
Mehdi Mirza and Simon Osindero.
\newblock Conditional generative adversarial nets.
\newblock {\em arXiv preprint arXiv:1411.1784}, 2014.

\bibitem{miyato2018spectral}
Takeru Miyato, Toshiki Kataoka, Masanori Koyama, and Yuichi Yoshida.
\newblock Spectral normalization for generative adversarial networks.
\newblock {\em arXiv preprint arXiv:1802.05957}, 2018.

\bibitem{nadeem2016computer}
Saad Nadeem and Arie Kaufman.
\newblock Computer-aided detection of polyps in optical colonoscopy images.
\newblock {\em SPIE Medical Imaging}, 9785:978525, 2016.

\bibitem{pajot2018unsupervised}
Arthur Pajot, Emmanuel de Bezenac, and Patrick Gallinari.
\newblock Unsupervised adversarial image reconstruction.
\newblock {\em International Conference on Learning Representations (ICLR)},
  2019.

\bibitem{porav2019reducing}
Horia Porav, Valentina Musat, and Paul Newman.
\newblock Reducing steganography in cycle-consistency gans.
\newblock {\em Proceedings of the IEEE Conference on Computer Vision and
  Pattern Recognition Workshops}, pages 78--82, 2019.

\bibitem{rau2019implicit}
Anita Rau, PJ~Eddie Edwards, Omer~F Ahmad, Paul Riordan, Mirek Janatka,
  Laurence~B Lovat, and Danail Stoyanov.
\newblock Implicit domain adaptation with conditional generative adversarial
  networks for depth prediction in endoscopy.
\newblock {\em International Journal of Computer Assisted Radiology and
  Surgery}, 14(7):1167--1176, 2019.

\bibitem{rex2017colorectal}
Douglas~K Rex, C~Richard Boland, Jason~A Dominitz, Francis~M Giardiello,
  David~A Johnson, Tonya Kaltenbach, Theodore~R Levin, David Lieberman, and
  Douglas~J Robertson.
\newblock Colorectal cancer screening: {R}ecommendations for physicians and
  patients from the {US Multi-Society Task Force on Colorectal Cancer}.
\newblock {\em The American Journal of Gastroenterology}, 112(7):1016, 2017.

\bibitem{schops2019surfelmeshing}
Thomas Sch{\"o}ps, Torsten Sattler, and Marc Pollefeys.
\newblock Surfelmeshing: Online surfel-based mesh reconstruction.
\newblock {\em IEEE Transactions on Pattern Analysis and Machine Intelligence},
  2019.

\bibitem{shin2018abnormal}
Younghak Shin, Hemin~Ali Qadir, and Ilangko Balasingham.
\newblock Abnormal colon polyp image synthesis using conditional adversarial
  networks for improved detection performance.
\newblock {\em IEEE Access}, 6:56007--56017, 2018.

\bibitem{whelan2015elasticfusion}
Thomas Whelan, Renato~F Salas-Moreno, Ben Glocker, Andrew~J Davison, and Stefan
  Leutenegger.
\newblock Elasticfusion: Real-time dense slam and light source estimation.
\newblock {\em The International Journal of Robotics Research},
  35(14):1697--1716, 2016.

\bibitem{yi2017dualgan}
Zili Yi, Hao Zhang, Ping Tan, and Minglun Gong.
\newblock {DualGAN}: Unsupervised dual learning for image-to-image translation.
\newblock {\em Proceedings of the IEEE International Conference on Computer
  Vision}, pages 2849--2857, 2017.

\bibitem{zhu2017unpaired}
Jun-Yan Zhu, Taesung Park, Phillip Isola, and Alexei~A Efros.
\newblock Unpaired image-to-image translation using cycle-consistent
  adversarial networks.
\newblock {\em Proceedings of the IEEE International Conference on Computer
  Vision}, pages 2223--2232, 2017.

\end{thebibliography}
\end{document}